\documentclass[usenatbib]{mn2e}
\usepackage[utf8]{inputenc}
\usepackage{natbib,epsfig,cite,mystyle,amsmath,amssymb,graphicx}
\usepackage{array, xcolor, caption}
\usepackage[colorlinks=true,linkcolor=blue,citecolor=blue]{hyperref}

\pdfoutput=1  % togliere il commento se le figure sono pdf, lasciarlo se sono eps
\usepackage{bm}
\usepackage{graphicx,color}
\usepackage{latexsym,amsmath,amssymb,graphicx,booktabs}
\usepackage{hyperref}
\usepackage{placeins}
\usepackage{subfigure}
\usepackage{mathtools}
\usepackage[normalem]{ulem}

\definecolor{MyBlue}{rgb}{0.15,0.15,0.70}
\definecolor{Dgreen}{rgb}{0,0.7,0.0}

\hypersetup{
colorlinks=true,
citecolor=MyBlue,
linkcolor=MyBlue,
urlcolor=MyBlue
}

\usepackage{amssymb}
\usepackage{amsmath}
\usepackage{amsfonts}
\usepackage{upgreek}
\usepackage{latexsym}
\usepackage{appendix}
\usepackage{eufrak}
\usepackage{dsfont}

\newcommand\ees{\end{eqnarray}}
\newcommand\bees{\begin{eqnarray}}

\usepackage[export]{adjustbox}

\newcommand\spart{\;\raise1.0pt\hbox{/}\hskip-6pt\partial}
\newcommand\spartb{\;\overline{\raise1.0pt\hbox{/}\hskip-6pt\partial}}

\newcommand{\be}{\begin{equation}}
\newcommand{\ee}{\end{equation}}

\newcommand{\beqa}{\begin{eqnarray}}
\newcommand{\eeqa}{\end{eqnarray}}

\newcommand{\nn}{\nonumber}

\begin{document}
\title[Characterisation of lensing selection effects for LISA ]{Characterisation of lensing selection effects for LISA massive black hole binary mergers}
\author[G. Cusin, N. Tamanini]{Giulia Cusin$^{1}$\thanks{giulia.cusin@unige.ch}, Nicola Tamanini$^{2}$ \\
$^{1}$Universit\'e de Gen\`eve, D\'epartement de Physique Th\'eorique and Centre for Astroparticle Physics,
24 quai Ernest-Ansermet, \\CH-1211 Gen\`eve 4, Switzerland \\
$^{2}$Max Planck Institute for Gravitational Physics (Albert Einstein Institute), Am M\"{u}hlenberg 1, Potsdam-Golm, 14476, Germany \\
}

\pagerange{\pageref{firstpage}--\pageref{lastpage}} \pubyear{2019}
\maketitle
\label{firstpage}

\begin{abstract}
We present a method to include lensing selection effects due to the finite horizon of a given detector when studying lensing of gravitational wave (GW) sources.
When selection effects are included, the mean of the magnification distribution is shifted from one to higher values for sufficiently high-redshift sources.
This introduces an irreducible (multiplicative) bias on the luminosity distance reconstruction, in addition to the typical source of uncertainty in the distance determination.
We apply this method to study lensing of GWs emitted by massive black hole binary mergers at high redshift detectable by LISA. We estimate the expected bias induced by selection effects on the luminosity distance reconstruction as function of cosmological redshift, and discuss its implications for cosmological and astrophysical analyses with LISA.  We also reconstruct the distribution of lensing magnification as a function of the observed luminosity distance to a source, that is the observable quantity in the absence of an electromagnetic counterpart. 
Lensing provides the dominant source of errors in distance measurements of high-redshift GW sources. Its full characterisation, including the impact of selection effects, is of paramount importance to correctly determine the astrophysical properties of the underlying source population and to be able to use gravitational wave sources as a new cosmological probe.
\end{abstract}


%%%%%%%%%%%%%%%%%%%%%%%%%%%%%%%%%%%%%%%%%%%%%%%%%%
\section{Introduction} 
%%%%%%%%%%%%%%%%%%%%%%%%%%%%%%%%%%%%%%%%%%%%%%%%%%

The \textit{Laser Interferometer Space Antenna} (LISA), an ESA-led space mission with launch expected for 2034, will open a new low-frequency window in the gravitational-wave (GW) spectrum \citep{Audley:2017drz}, which should lead to the detection of a plethora of new GW sources promising to double the recent revolutionary discoveries achieved at high frequencies by the LIGO-Virgo interferometers \citep{Abbott:2016blz,Abbott:2016nmj,TheLIGOScientific:2016pea,Abbott:2017gyy,Abbott:2017oio,Abbott:2017vtc,TheLIGOScientific:2017qsa,LIGOScientific:2018mvr,Abbott:2020niy}.
Among all the different GW sources that LISA will observe, the merger of massive black hole binaries (MBHB) in the $10^4 - 10^7$ solar mass range is considered one of the key targets of the mission \citep{Klein:2015hvg}.
Their detection will allow us to probe for the first time the true nature of the massive dark objects found at the centers of galaxies, testing in this way the hypothesis that MBHBs come from the inevitable outcome of galaxy assembly, and consequently unveiling the origin and growth history of MBHs at high redshift \citep{Haehnelt:1994wt,Kauffmann:1999ce,2003ApJ...582..559V,Enoki:2004ew,Sesana:2004gf,Micic:2007vd,Barausse:2012fy,Colpi:2014poa,Umeda:2016smj,2016MNRAS.457.3356V}.

MBHB mergers will be detected by LISA up to very high redshift \citep{Klein:2015hvg}, possibly up to $z\sim 20$ if any of them can be formed so early in the cosmic history.
Intervening matter along the line of sight between the source and the observer will lens the GW signal, changing some of the properties from the emitted to the observed waveform, in particular magnifying/de-magnifying its amplitude and consequently shifting the measured value of the luminosity distance, hence biasing the inferred source-frame chirp mass\footnote{From the waveform one can extract the redshifted chirp mass of the source. If the distance reconstruction is biased, also redshift associated to that distance (by fixing a reference cosmological model) is biased, hence the reconstruction of the source-frame chirp mass is biased as well.}.
These effects will be particularly relevant for the few LISA MBHB events that will appear strongly lensed \citep{Sereno:2010dr,Sereno:2011ty}.

In this study we present a method to estimate the effect of lensing on the luminosity distance of high redshift GW astrophysical sources, such as LISA MBHBs, including for the first time selection effects due to the finite horizon of the GW detector. 
%In this study we will assess the contribution of lensing due to cosmic structures on the LISA measurements of MBHBs, taking into account for the first time selection effects due to the finite sensitivity of the instrument.
The impact of gravitational lensing on the distance measurement for MBHBs was already pointed out by \citet{Holz:2005df, Hilbert:2010am, Hirata:2010ba, Shang_2010}, where it is estimated that lensing will introduce an insurmountable error of a few percent on the distance estimate, for each individual high redshift binary system. 
In these works, the effect of lensing is directly estimated from the distribution of lensing magnification as a function of cosmological redshift, typically obtained from ray-tracing simulations (e.g. \citet{Sato_2009, Takahashi:2011qd}).
However a realistic GW detector has a finite sensitivity: magnified sources are on average easier to detect than de-magnified ones and this affects the distribution of lensing magnification of an observed source sample.
These effects, hereafter dubbed \emph{selection effects}, are usually neglected when estimating the lensing-induced uncertainty on the cosmological distance measurement from high-redshift GW sources. 
%In Ref.\,\citet{Bonvin:2016qxr}, line of sight effects on the waveform for BBH are studied and it is shown that weak lensing affects the waveform in a way that is totally subdominant with respect to the effect given by peculiar velocities and accelerations and does not significantly affect the reconstruction of the intrinsic parameters of the emitting source. 
 
When selection effects are included, the mean of the magnification distribution is shifted from 1 as we go farther in redshift. This introduces an irreducible (multiplicative) bias on the distance reconstruction,  independent of the sample size.
We apply this method to estimate the effect of lensing on a population of MBHB mergers detectable by LISA and we compute bias and uncertainty of the distance estimator as functions of cosmological redshift. 
We also compute the distribution of lensing magnification as a function of the observed luminosity distance (using the entire source population), that is the observable quantity in the absence of an EM counterpart.

Our results have direct impact on cosmological and astrophysical investigations made with LISA.
MBHBs mergers can indeed be used as \textit{standard sirens} \citep{Schutz:1986gp} to probe the rate of expansion of the universe at high-redshift whenever an electromagnetic (EM) counterpart can be identified.
Recent forecasts show that LISA, in combination with suitable EM facilities, will provide few MBHB standard sirens per year up to $z\simeq 8$ \citep{Tamanini:2016zlh}, yielding stringent constraints at high redshift not only on $\Lambda$CDM but also on possible deviations from it \citep{Caprini:2016qxs,Cai:2017yww,Belgacem:2019pkk,Calcagni:2019kzo,Calcagni:2019ngc,Corman:2020pyr,Speri:2020hwc}.
On the other hand, high-redshift MBHB mergers can also be used to test the formation, evolution and environment of MBHs \citep{Sesana:2007sh,Sesana:2010wy,Colpi:2019yzd,Mukherjee_2020}, and thus shed light on the astrophysical properties of their population across cosmic time.
Here we show how lensing selection effects affect these investigations, discussing in particular the impact that our results have on the cosmological and astrophysical analyses that have been proposed for LISA.

In what follows we set $c=1$, but keep the gravitational constant $G$. We assume vanishing spatial curvature and we set the present scale factor, $a_0=1$. We set $\Lambda$CDM as our fiducial cosmology with cosmological parameters taken from \citet{Takahashi:2011qd}.\\

  %\section{General method}\label{sec:method}
  %\subsection{Inclusion of selection effects}\label{e:stat}
\section{Inclusion of selection effects} 

We follow the derivation of \citet{Cusin:2019eyv}.  
A realistic GW detector has a limiting strain sensitivity, usually cast as an SNR limit, $\rho_{\lim}$. A source is detected if $\rho\geq\rho_{\lim}$, where the SNR $\rho$ is defined by~\citep{Finn:1992xs}
\be
\rho^2 = 4\int \frac{|h(f)|^2}{S_n(f)}df  =\int \rho^2(f)df\,.
\label{eq:SNR_def}
\ee
Here $S_n(f)$ is the noise of the detector at frequency $f$, see e.g.~\citet{Maggiore:1900zz}. 

If an event is magnified by $\mu>1$ we could see it even if without magnification its SNR is below threshold.
In other terms, denoting with $\rho_{\lim}$ the SNR threshold for detection in the absence of lensing, in general in the presence of magnification $\mu>1$, the threshold for detection lowers to $\rho_{\lim}/\sqrt{\mu}$.
%<\rho_{\lim}$. 
It follows that
in the presence of magnification,
the total number of objects which we expect to see from a redshift  bin $dz$ is given by the following convolution %\giulia{We removed the $dz$ from the equation below}
\citep{Cusin:2019eyv}
\be\label{Nobs1}
d\mathcal{N}_{\rm obs}(z, \rho_{\lim}) =\int_0^\infty \!\!d\mu p(\mu,z)d\mathcal{N}(\mu, z) \,,
\ee
where $p(\mu,z)$ is the distribution of lensing magnification as a function of cosmological redshift and 
\be\label{ForPlot1}
d\mathcal{N}(\mu, z)\equiv dz\int_{\rho_{\lim}/\sqrt{\mu}}^\infty N(z, \rho)d\rho \,,
\ee
is the number of sources that we see from  redshift  $z$ in the presence of magnification $\mu$ and $N(z, \rho)$ is the number density of sources as function of redshift and SNR $\rho$.\footnote{In Eq.\,(\ref{ForPlot1}) we have denoted with $\rho$ the SNR of a source computed assuming no lensing effects.} The mean amplification of a source at redshift $z$ is
\be\label{ampl1}
\langle\mu\rangle(z) = \frac{\int_0^\infty \!\!d\mu\, \mu \,p(\mu,z)d\mathcal{N}(\mu, z)}{\int_0^\infty \!\!d\mu\, p(\mu,z)d\mathcal{N}(\mu, z)} \,.
\ee
Note that (\ref{ampl1}) is the first moment of the probability distribution of magnification as seen by the detector (i.e.~with selection effects included) 
\be\label{PDistr}
\mathcal{P}_z(\mu,z)\equiv \mathcal{C}\, p(\mu,z)\frac{d\mathcal{N}(\mu, z)}{dz}\,,
\ee
where $\mathcal{C}$ is a  normalization constant. Eq.\,(\ref{PDistr}) depends on three basic ingredients: \emph{i)} distribution of magnification $p(\mu, z)$, \emph{ii)} distribution of sources as a function of redshift and luminosity (or analogously $\rho$) \emph{iii)} sensitivity curve of a given experiment. 
%For sufficiently high $\mu$, $p(\mu,z) \simeq - d\tau/d\mu\propto \mu^{-3}$, while the number of sources which we see with magnification $\mu$ or more from a redshift bin around $z$, defined in eq.\,(\ref{ForPlot1}), 

%The functions determined here, however are not observable as, 
In the absence of an EM counterpart, we do not know the cosmological redshift of a GW event. We usually infer it by assuming that the observed 
luminosity distance is the one of the background universe without magnification. In the presence of magnification, the relation between observed and cosmological luminosity distance is given by $D_{\text{obs}}(z,\mu)\equiv D(z)/\sqrt{\mu}$.  
One can rewrite (\ref{Nobs1}) and (\ref{ampl1}) as functions of the observed luminosity distance extracted from observations, using their relation in terms of magnification. Then in full analogy with eq.\,(\ref{PDistr}), the probability distribution of magnification at the detector, as a function of the observed luminosity distance of the source, is given by \citep{Cusin:2019eyv}
\be\label{PDistr2}
\mathcal{P}_{\text{obs}}(\mu, D_{\text{obs}})\equiv \mathcal{C}\, p(\mu, D_{\text{obs}})\frac{d\mathcal{N}(\mu, D_{\text{obs}})}{dD_{\text{obs}}}\,, 
\ee
where $\mathcal{C}$ is a constant of normalization and $d\mathcal{N}(\mu, D_{\text{obs}})/dD_{\text{obs}}$ is the number of sources observable at observed distance $D_{\text{obs}}$ if the magnification is $\mu$, explicitly given by \citep{Cusin:2019eyv}
\be\frac{d\mathcal{N}(\mu, D_{\text{obs}})}{dD_{\text{obs}}}=\frac{\sqrt{\mu}}{D'(z(D_{\text{obs}}, \mu))}\int_{\rho_{\text{lim}}/\sqrt{\mu}}^{\infty} N(\rho, z)d\rho\,,
\ee
where $z=z(D_{\text{obs}}, \mu)$ and a prime denotes a derivative with respect to the argument.\\
%\giulia{Added definition above to make JP less lost} \\

%\subsection{Effects on Hubble diagram}\label{estimator}

\section{Luminosity distance estimator}\label{estimator}

Let us assume that we observe a sample of sources with EM counterpart. From each observed event we extract a luminosity distance $D_{\text{obs}}$ and redshift $z$, and for each source, the observed and cosmological distance are related through $D^2=\mu D_{\text{obs}}^2$. We bin sources in redshift and we associate to each bin $z$ an average \emph{observed} luminosity distance as
\be
\bar{D}^2_{\text{obs}}\equiv \left[\frac{1}{N(z)}\sum_{i=1}^{N} D^2_{\text{obs},i}\right]\,,
\ee
where $N(z)$ is the number of sources in the bin $z$. 

We write a (formal) expression for the true (unlensed) luminosity distance at a given redshift as
\be\label{defD}
\hat{D}^2(z)=\frac{1}{N(z)}\sum_{i=1}^{N}\mu_i D_{\text{obs}, i}^2 \simeq \mu(z) \bar{D}^2_{\text{obs}}\,. 
\ee
To go to the last equality in eq.\,(\ref{defD}) we used the fact that the magnification of a single source is not observable and we replaced $\mu_i\rightarrow \mu(z)$ where $\mu(z)$ is a stochastic parameter which follows the distribution $\mathcal{P}_z(\mu, z)$. 
We stress that (\ref{defD}) has to be interpreted as a formal expression: only its average and higher moments  can be computed from data and theoretical inputs such as the PDF of magnification computed from ray-tracing simulations.
Denoting with $\langle \dots \rangle$ the average of a function of $\mu$ computed with the distribution $\mathcal{P}_z(\mu, z)$, we have that 
\be\label{mean}
\langle \hat{D}^2(z)\rangle=\langle \mu (z)\rangle\bar{D}^2_{\text{obs}}\,,
\ee
where $\langle \mu (z)\rangle$ is explicitly given by \eqref{ampl1}. We observe that the quantity $\langle \mu(z)\rangle$ is traditionally assumed to be one, see e.g. \citet{Hirata_2010, Holz:2004xx}.  However, once selection effects due to the sensitivity curve of a given observatory are included, flux conservation is not enforced anymore and in general $\langle \mu (z)\rangle\neq 1$, hence this amplification factor needs to be added to get an unbiased estimate of the cosmological distance.  We also stress that in eq.\,(\ref{mean}) two independent average procedures are involved: an arithmetic mean to compute the average distance of the sample and a stochastic mean with the magnification PDF.\footnote{Note that the actual bias on the distance is given by $\langle \mu^{1/2}(z)\rangle$. We decide however to work with a quadratic quantity in eq.\,(\ref{defD}) to stress the effect of lensing selection, as the bias on the distance square is identically one if selection effects are not accounted for.}

To estimate the uncertainty on the determination of the cosmological distance, we follow \citet{Hirata_2010} and we compute a variance given by the inverse of the Fisher information. We get (see appendix for details) 
\be\label{sigma}
\sigma(\ln \hat{D}^2)\simeq \frac{1}{\sqrt{I(x)}}\,,
\ee
where $x\equiv \ln \mu$ and $I(x)$ is the Fisher information
\be\label{I}
I(x)\equiv N(z) \int \mathcal{P}_z(x, z)\left[\frac{d}{dx}\ln \mathcal{P}_z(x, z)\right]^2 dx\,. 
\ee
Notice that while in \citet{Hirata_2010} the Fisher information is computed from the distribution of magnification $p(x, z)$ which depends only on the matter distribution of the Universe, we compute the Fisher information from the distribution, 
$\mathcal{P}_z(x, z)$,  which contains selection effects specific to a given observatory.

The reason why we estimate the variance of (\ref{defD}) using its Cramer-Rao bound instead of computing the second moment of the estimator with the PDF of $\mu$ is that  the latter, namely $\sigma_{\mu}$, is a divergent quantity: the PDF decays as $\mu^{-3}$ at high $\mu$, hence $\sigma_{\mu}$ has an ultraviolet logarithmic divergence. However, this divergence is not physical and it is a consequence of the fact that, to predict the high-magnification tail of the magnification distribution, we are artificially assuming that ray optics is a good approximation at any scale. Realistically, wave effects appear below a given frequency-dependent scale and regularise this divergence effectively introducing a frequency-dependent ultraviolet cut-off in $\mu$. To describe this effect properly, one should re-run a ray-tracing simulation introducing diffraction effects below a given frequency dependent scale.
Such an investigation is outside the scope of the present study. Instead of introducing artificial cut-offs in the computation of the the second moment of the magnification distribution, we decided here to estimate the variance of the estimator using its Cramer-Rao bound.\footnote{We checked that estimating the variance directly from (\ref{defD}) adding by hand an upper cut-off in $\mu$ gives results in qualitative agreement with the Cramer-Rao estimate (e.g. within $8\%$ if we artificially fix this cut-off at $\mu=5$).}

In the absence of an EM counterpart, we have access only to the observed luminosity distance $D_{\text{obs}}$ of a source. If we have a set of $N$ sources with observed distances distributed in a bin with mean $D_{\text{obs}}$, the average (true, cosmological) luminosity distance associated to the sample is given by
\be\label{DDobs}
\langle\hat{D}^2(D_{\text{obs}})\rangle=\langle \mu(D_{\text{obs}})\rangle D^2_{\text{obs}}\,,
\ee
where  $\langle \mu(D_{\text{obs}})\rangle$ is the mean magnification associated to the distribution of magnification $\mathcal{P}_{\text{obs}}$ as function of the observed distance defined in eq.\,(\ref{PDistr2}). Eq.\,(\ref{DDobs}) gives an estimate of the (real) luminosity distance of a sample as a function of the observed luminosity distance, and corresponds to eq.\,(\ref{mean}), computed as function of the sample observed distance (instead of redshift).  
The uncertainty associated to the estimator (\ref{DDobs}) is given by eq.~(\ref{sigma}) where the Fisher information (\ref{I}) has to be computed using the probability distribution (\ref{PDistr2}).

In the rest of this paper we will often refer to the average magnification $\langle \mu \rangle \neq 1$ as bias on distance estimator, understanding that if the presence of magnification is not accounted for, we get a biased reconstruction of a source luminosity distance. We stress that this is somehow an abuse of language: the estimator formally defined in (\ref{defD}) is actually unbiased by construction.

\begin{figure}
\begin{center}
\includegraphics[width=0.47\textwidth]{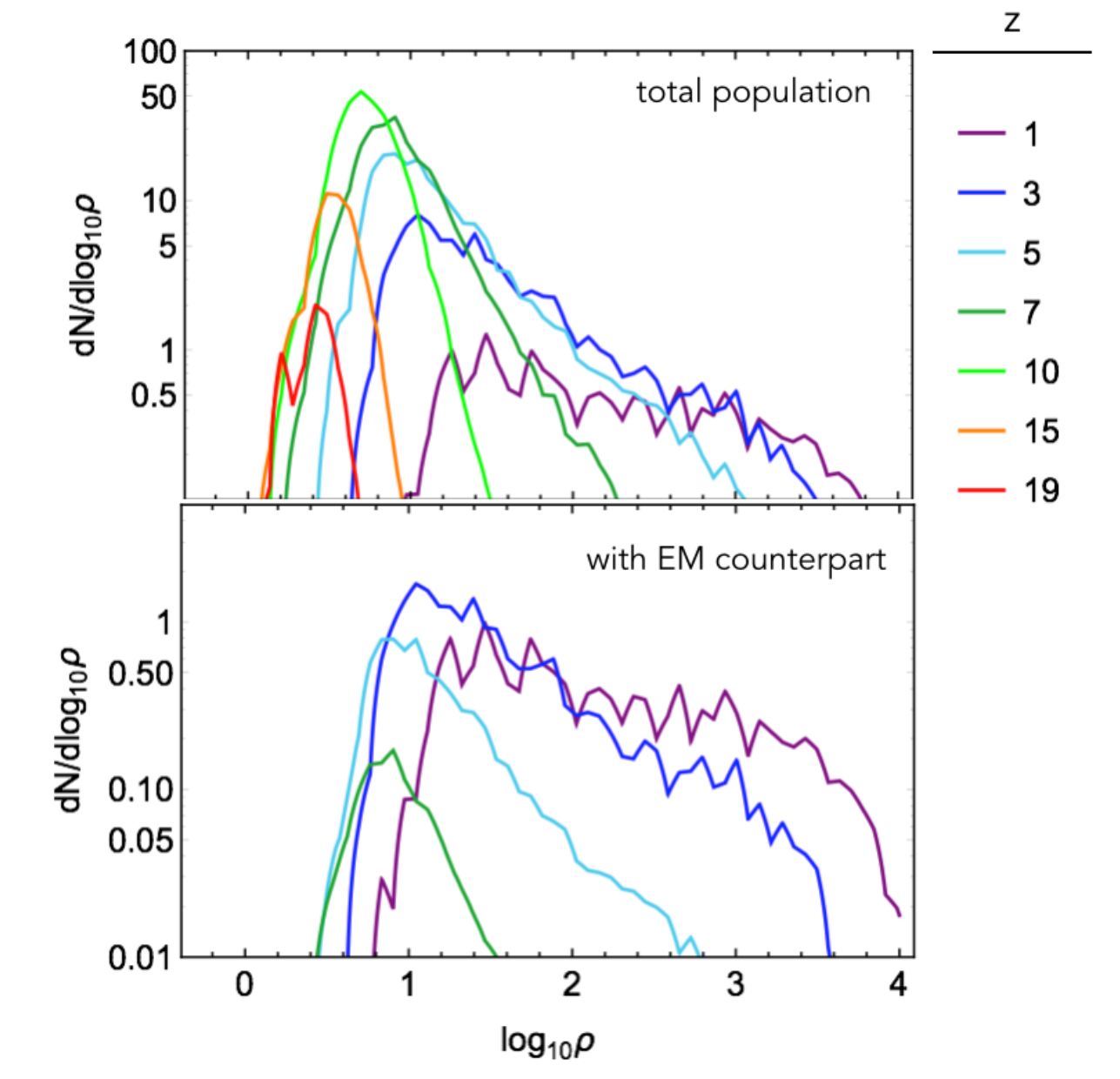}
\end{center}
\caption{\label{pop} Number of events per unit redshift, for the total population and the sub-population with EM counterpart.  Each line corresponds to the number of events in a redshift  bin   $z \pm 0.5$ and the values of $z$ are listed in the legend on the right of the figure. We fixed an observation time  $T_{\text{obs}}=1$ year. The number of events increase linearly with $T_{\text{obs}}$. }
\end{figure}

%\section{Results for LISA}\label{s:num}
\section{Results for LISA} 

We apply the analysis presented in the previous section to the case of MBHBs detectable by LISA. 
Our first goal is to compute the bias $\langle \mu^{1/2}(z)\rangle$ and the uncertainty of the distance estimator $\hat{D}$, for the full population of MBHBs and the sub-population with EM counterpart.
%We also compute the distribution of magnification for the total MBHB population  (with or without EM counterpart), including selection effects due to the finite sensitivity of LISA.
Three ingredients enter the computation: 
%\begin{enumerate}[label=(\roman*)]
  %  \item 
(i) The distribution of magnification as a function of cosmological redshift $p(\mu, z)$, which we take from the high-resolution ray-tracing simulations of \citet{Takahashi:2011qd}. The box size of the simulation is 50$h^{-1}$Mpc with $1024^3$ particles, the mean particle separation is 50$h^{-1}$kpc, and the softening length of $h^{-1}$kpc. The resolution is such that the simulation predicts both the low magnification contribution to the PDF due to the large scale structure, and the strong lensing high-magnification tail up to halo scales, which is particularly relevant for our analysis. The simulation does not consider the effect of baryons, which typically enhance the strong magnification tail. This means that the results we will get using the simulations of \citet{Takahashi:2011qd} for lensing selection effects are rather conservative. We also observe that GW in the LISA band are insensitive to sub-galactic structures as diffraction becomes very effective on those scales \citep{Takahashi:2003ix}. %rendering those lenses \emph{transparent} to the wave. 
(ii) The LISA strain noise curve, which we take from \citet{Audley:2017drz,Robson_2019}. We include the contribution from galactic binaries to the noise curve, assuming a nominal duration of the mission of 4 years.
To estimate the SNR in eq.~\eqref{eq:SNR_def}, we employ the phenomenological inspiral-merger-ringdown waveform model known as PhenomA \citep{Ajith_2007}.
While more accurate models now certainly exist, which include spin-precession and other effects, for our purposes PhenomA is sufficiently accurate to estimate the SNR, and in fact gives results compatible with the ones of \citet{Klein:2015hvg,Tamanini:2016uin} for the number of events above threshold. 
(iii) The astrophysical model for MBHB distribution. We use here for definiteness the catalogues used in \citet{Klein:2015hvg}, based on the MBHB populations of \citet{Barausse:2012fy}.
Note that although there are more recent studies yielding updated populations of MBHBs for LISA (e.g.~\citet{Salcido:2016oor,Bonetti:2018tpf,Katz:2019qlu,2019MNRAS.486.2336D}), for our case study we choose to work with the catalogues used in \citet{Klein:2015hvg} since for these we can also retrieve the fraction of the population with EM counterpart from \citet{Tamanini:2016zlh,Tamanini:2016uin}.
We focus in particular on the ``popIII'' population model based on light growth seeds for MBHs, since it is more affected by lensing selection effects, as we will show below.
In what follows all results will refer to this model, unless otherwise specified.
We provide in fact results for the other two ``Q3'' heavy-seed MBH population models considered in \citet{Klein:2015hvg} whenever they are relevant, and review their cases in more details in the discussion below.
%provides larger  average results with respect to all populations models considered in \citet{Klein:2015hvg}.
% (for the same astrophysical model).  
%To determine the fraction and redshift distribution of MBHB with detected EM counterpart, we use results from \citet{Tamanini:2016zlh,Tamanini:2016uin}.
The distribution of MBHB mergers in the popIII model as a function of SNR and redshift is presented in Fig.~\ref{pop}, for both the full population and the sub-population with EM counterpart.
    %\nt{Is this enough or should I add more details?}
%\end{enumerate}

\begin{figure}
\begin{center}
\includegraphics[width=0.46\textwidth]{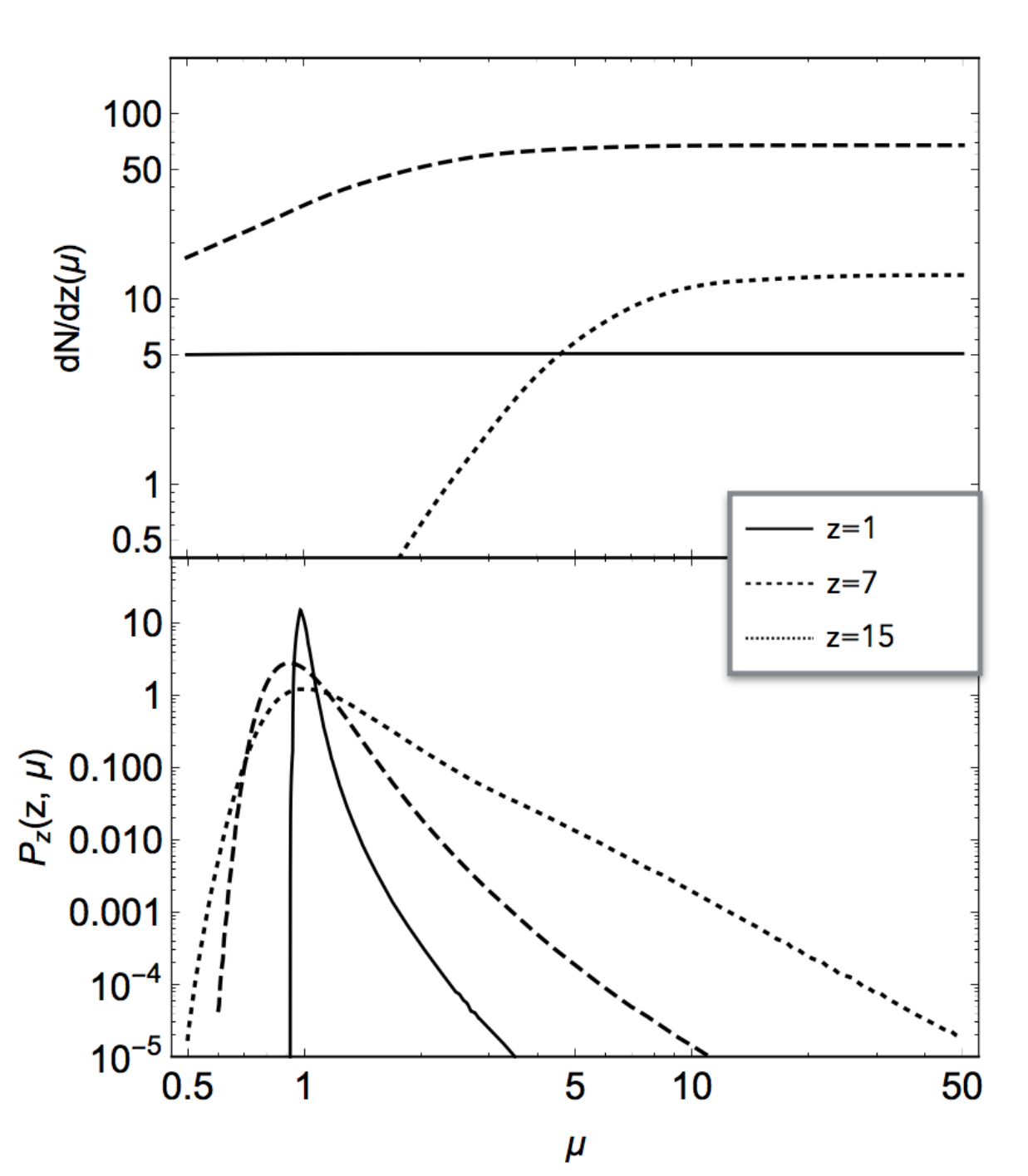}
\end{center}
\caption{\label{NP}Top: Number of sources from a given redshift, visible if the magnification is $\mu$ as a function of magnification, see eq.\,\eqref{ForPlot1}. We assumed $T_{\text{obs}}=4$ yr. Bottom: Distribution of magnification (cf.~eq.\,(\ref{ampl1})) with selection effects included. Each line is obtained multiplying the distribution $p(\mu, z)$ of \citet{Takahashi:2011qd} by the corresponding  $dN/dz$ in the top panel (and normalising).  }
\end{figure}

The distribution of magnification for LISA, once selection effects are included, eq.\,(\ref{ampl1}),  is shown in the bottom panel of Fig.\,\ref{NP}. Each line corresponds to a cosmological redshift and it is obtained multiplying the distribution of magnification of the ray-tracing simulation of \citet{Takahashi:2011qd} by the number of objects visible with LISA from a given redshift, as a function of magnification, eq.\,(\ref{ForPlot1}). This latter is shown in the top panel of Fig.\,\ref{NP}, for different redshifts. The number of events visible from a given redshift is a monotonically increasing function of magnification, and it reaches a plateau at the value of magnification for which all sources from that redshift have been observed. As expected, the value of magnification at which the plateau is reached is a monotonically increasing function of redshift. It follows that selection effects are more important for high redshift sources than for low redshift ones.
%Selection effects for sources at $z=12$ is shown in Fig.\,\ref{PComp}.\giulia{say that they decrease with redshift. Give fractional differences?} 
%\subsection{Distribution of magnification for LISA}

\begin{figure}
\begin{center}
\includegraphics[width=0.46\textwidth]{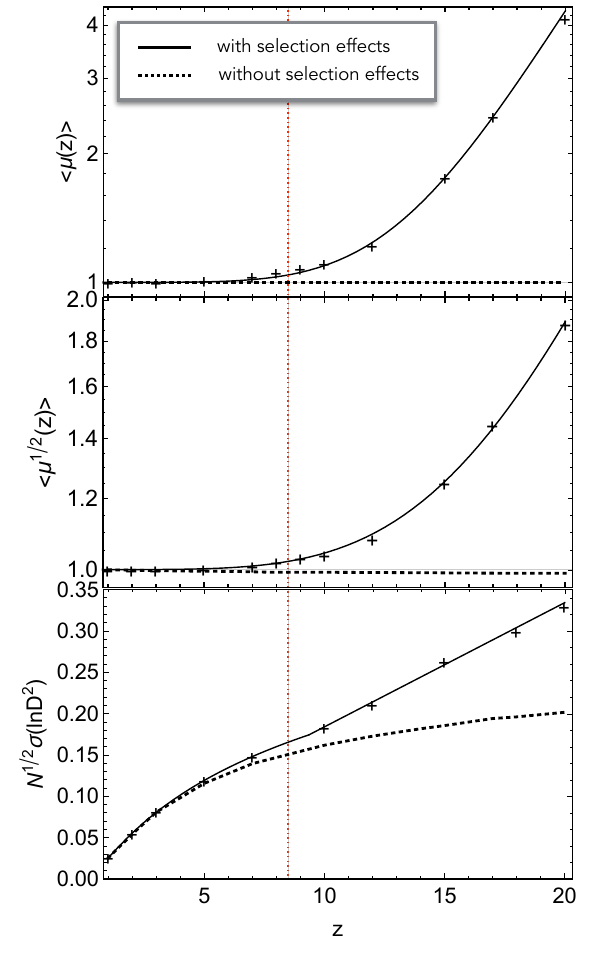}
\end{center}
\caption{\label{FinalPlot}
Bias and uncertainty of the distance estimator $\hat{D}$ as a function of cosmological redshift. We show both the bias on distance and on distance square, $\langle \mu^{1/2}\rangle$ and $\langle \mu\rangle$, respectively. We explicitly show the effect of including selection effects in the analysis: $\langle \mu\rangle$ is identically one if selection effects are not included as a consequence of flux conservation. The solid lines are the fitting formulae in eqs.~\eqref{eq:error_fit} and \eqref{eq:bias_fit}. With a vertical red line we indicate the maximum redshift at which for our astrophysical model, an EM counterpart is observable. 
}
\end{figure}

Finally we compute the bias  and the uncertainty of the distance estimator $\hat{D}$. We observe that the bias is independent of the sample size while the variance scales as $\propto 1/\sqrt{N}$ for a sufficiently large sample. In Fig.\,\ref{FinalPlot}, %\nt{Fig.~\ref{FinalPlot} is mentioned before Fig.~\ref{NPobs}}\giulia{I've exhanged the two figures so that they ar in order}
we compare the results with what one would obtain neglecting selection effects: for the variance one finds a low-redshift behaviour similar to the one reported in \citet{Hirata_2010}, while at high redshift selection effects become important.
Likewise, at low redshift the bias is %identically one
irrelevant as flux is conserved when selection effects are negligible, while above $z\simeq 8$, corresponding to the peak of the MBHB population distribution, the bias becomes strongly marked.

%\nt{try also to explain expected differences between Holz vs Takahashi}\\
%\giulia{It would be good to give a fit}\nt{I'll add the fit here if not given somewhere else}\\ 

By fitting the numerical curves we obtained in Fig.~\ref{FinalPlot}, we can provide a simple analytical estimate of the lensing uncertainty (1$\sigma$ deviation) as a function of redshift, in analogy to the formula provided by \citet{Hirata_2010} (eq.~(20)) but now containing selection effects.
We find that the following expression well fit the numerical data
%(AIC/BIC\footnote{Akaike Information Criterion/Bayesian Information Criterion.}$\sim -70$)
%\nt{I eliminated references to AIC/BIC as they are not necessary and can create confusion. If the referee will ask some numbers we will provide them later.}
at all redshift (for one siren $N=1$):
\begin{multline}
\sigma(\ln D^2)_{\rm fit}^{\rm popIII} = %2\left(\frac{\sigma(D)}{D}\right)^{\rm popIII}\nn\\
2\left.\frac{\sigma(D)}{D}\right|^{\rm popIII}_{\rm fit}\\
=\begin{cases}
  0.061 \left(\frac{1-(z+1)^{-0.264}}{0.264}\right)^{1.89} & \text{for }z\le 9.35\,,\\    
  0.034 + 0.015\, z & \text{for }z> 9.35\,.
\end{cases}
\label{eq:error_fit}
\end{multline}
%\giulia{added def, maybe we shoud improve formatting}\nt{what about this? Not great, but it's fine for me}
All numbers appearing in this formula have been used as parameters for the fit, including the cut in redshift.
Note that we used the same expression originally proposed in \citet{Hirata_2010} to fit the first part of the curve, finding parameter values close to the ones reported therein with a few percent global shift towards lower uncertainties which we associate to differences in the simulations of the lensing distribution between \citet{Hirata_2010} and \citet{Takahashi:2011qd}.\footnote{In particular, \citet{Takahashi:2011qd} derives the PDF of magnification using a high-resolution ray-tracing numerical simulation, while \citet{Hirata_2010} employs the PDF derived in \citet{Holz:1997ic} in the context of a stochastic universe method where the geodesic deviation equation is solved backward in time with an approach similar to that used in Swiss cheese universe models.}
At redshift higher than $z=9.35$ we used instead a simple linear fit capturing the behaviour due to selection effects.
The total fit given in eq.~\eqref{eq:error_fit} is shown in the bottom panel of Fig.~\ref{FinalPlot}.

We can furthermore derive for the first time a simple analytical fit of the expected bias due to lensing on distance measurements of MBHB mergers observed by LISA.
This is well fitted
%(AIC/BIC $\sim -90$) \giulia{give intuition of what this indicator is }
by the following simple polynomial function at all redshift (cf.~Fig.~\ref{FinalPlot}):
%This is well fitted (AIC/BIC $\sim -60$) by the following simple polynomial function at all redshift (cf.~Fig.~\ref{FinalPlot}):
\begin{equation}
    \left< \sqrt{\mu(z)} \right>_{\rm fit}^{\rm popIII} = 1 + 1.89 \times 10^{-6}\, z^{4.36} \,.
    \label{eq:bias_fit}
\end{equation}
For completeness we provide here also the fit of the bias on distance square (cf. eq.\,(\ref{mean})), which is identically 1 if lensing selection effects are not included:
%(AIC/BIC $\sim -60$): 
\begin{equation}
    \left< \mu(z) \right>_{\rm fit}^{\rm popIII} = 1 + 7.125 \times 10^{-7}\, z^{5.124} \,.
%\label{eq:bias_fit}
\end{equation}
By comparing the two expressions in eqs.~\eqref{eq:error_fit} and \eqref{eq:bias_fit} (for $N=1$), we can find the redshift above which the lensing bias will be more significant than the lensing uncertainty (at 1$\sigma$), and thus no longer negligible for distance measurements.
This happens at $z\simeq 15$. %$z\simeq 11.7$.\giulia{still the same this? I would expect it to change a bit}
Above this redshift, the LISA distance measurement of a single MBHB merger will on average expected to be biased by lensing effects, and thus not reliable.
%\nt{updated number and improved last sentence}

As the results above apply to the popIII population model only, we provide here also the fit equivalent to eq.~\eqref{eq:error_fit} for the heavy-seed Q3 populations of \citet{Klein:2015hvg,Barausse:2012fy}.
In such case selection effects are negligible at all redshift (cf.~discussion below), and we find that an optimal fit
%(AIC/BIC $\sim -140$)
is given by
\begin{align}
    \sigma(\ln D^2)_{\rm fit}^{\rm Q3} = 2\left.\frac{\sigma(D)}{D}\right|^{\rm Q3}_{\rm fit}= 0.096\, \left(\frac{1-(z+1)^{-0.62}}{0.62}\right)^{2.36} \,.
\end{align} 
This is nothing but the functional form proposed by \citet{Hirata_2010} fitted to the whole redshift range up to $z=20$, and constitutes an excellent representation of the numerical curve presented in the bottom panel of Fig.~\ref{FinalPlot} (dottet line).
Being selection effects negligible for heavy-seed MBHB population models, no relevant bias will appear.
In other words both $\left< \mu(z) \right>$ and $\left< \sqrt{\mu(z)} \right>$ are basically one at all redshift probed by LISA, on the contrary to what we found above for the popIII model (cf.~eq.~\eqref{eq:bias_fit}).

\begin{figure}
\begin{center}
\includegraphics[width=0.45\textwidth]{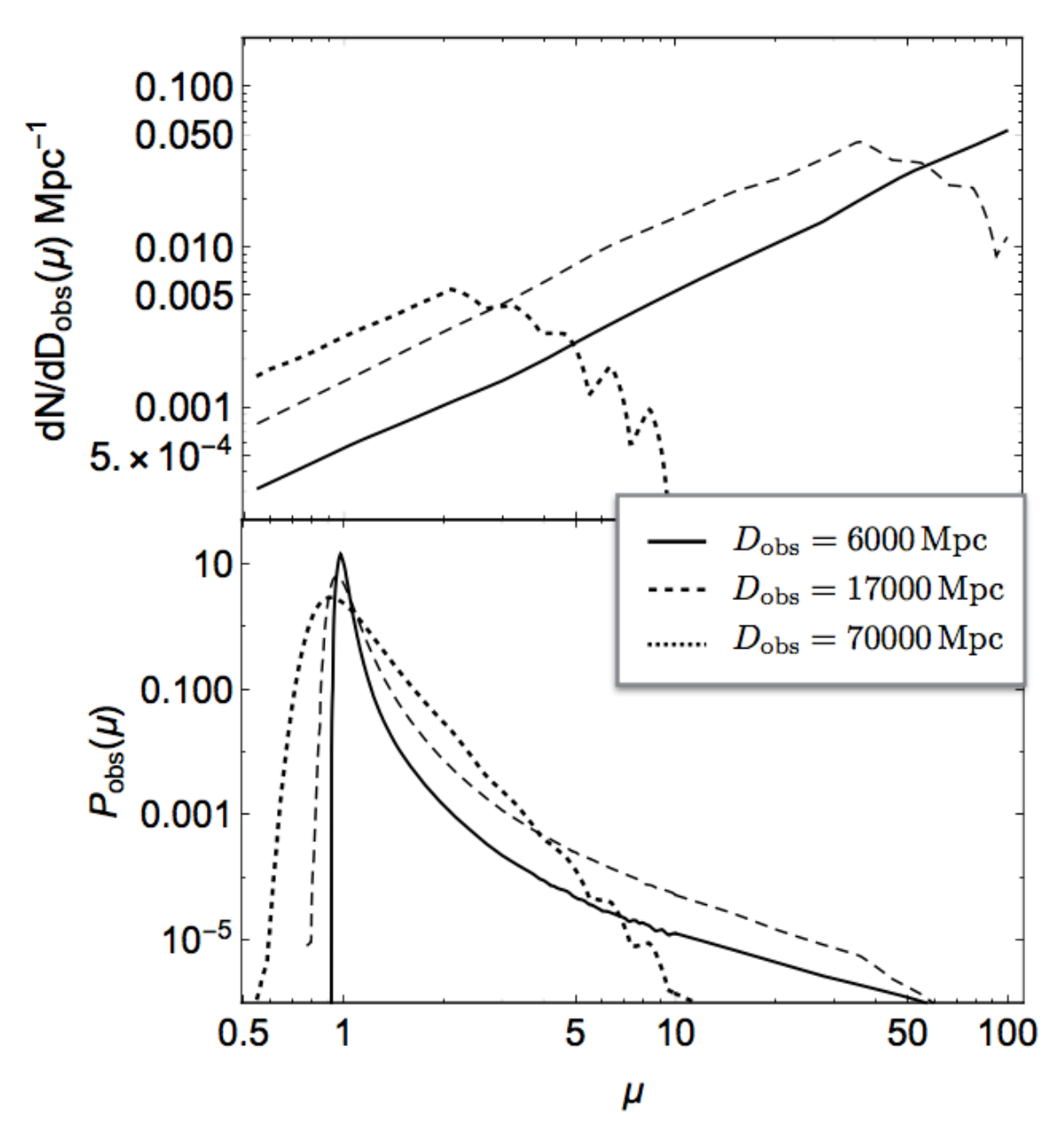}
\end{center}
\caption{\label{NPobs} Top: Number of sources from a given observed luminosity distance, visible if the magnification is $\mu$ as a function of magnification (cf.~eq.\,(\ref{PDistr2})). We assumed $T_{\text{obs}}=4$ yr. Bottom: Distribution of magnification (\ref{ampl1}) with selection effects included. Each line is obtained multiplying the distribution $p(\mu, D_{\text{obs}})$ computed from the simulation of \citet{Takahashi:2011qd} by the corresponding  $dN/dD_{\text{obs}}$ in the top panel (and normalising).
\label{obs_mu}
}
\end{figure}

Finally we provide results in terms of observable quantities.
In the absence of an EM counterpart, a direct access to the redshift of a GW source is not possible. The quantity that is directly observable in this case is the \emph{observed} luminosity distance of a given source (which can be converted into redshift only by assuming a cosmological model).
We compute the distribution of magnification with selection effects as a function of the observed luminosity distance, eq.\,(\ref{PDistr2}); see also  \citet{Cusin:2019eyv} for details on this derivation. In Fig.\,\ref{NPobs}  we plot the number of sources from a given observed luminosity distance, visible if the magnification is $\mu$ as a function of magnification. In the bottom panel of the same figure we show the distribution of the observed magnification  $\mathcal{P}_{\text{obs}}(\mu, D_{\text{obs}})$ obtained multiplying each line of the top panel by the PDF $p(\mu, D_{\text{obs}})$ obtained from \citet{Takahashi:2011qd} converting redshift to observed distance trough  $D(z)=\sqrt{\mu}D_{\text{obs}}$.

Comparing the top panels of Figs.\,\ref{NP}  and \ref{NPobs} it becomes apparent the role played by selection effects. The probability distribution of magnification  increases as we increase the cosmological redshift of the source (for magnification $\mu$ fixed). This is a consequence of the fact that the number of sources visible from a given cosmological redshift is a monotonically increasing function of magnification.  On the other hand, the number of sources visible from a given observed  distance if the magnification is $\mu$, for a fixed $D_{\text{obs}}$ has a more involved behavior. As we vary magnification, a given bin around $D_{\text{obs}}$ receives contributions from sources at different cosmological distances $D(z)=\sqrt{\mu} D_{\text{obs}}$. The peak of the distribution of magnification becomes broader as we increase the observed distance of a source (see bottom panel of Fig.\,\ref{NPobs}), but the high-magnification tail gets suppressed. Indeed if we observe a source at very high distance, it is improbable that this is a magnified one as it would come from an even higher cosmological distance $D(z)>D_{\text{obs}}$. 
%\giulia{I have added two paragraphs above}\nt{sounds good, but I rewrote the following sentence and added a comment at the end of the following paragraph}

\begin{figure}
\begin{center}
\includegraphics[width=0.45\textwidth]{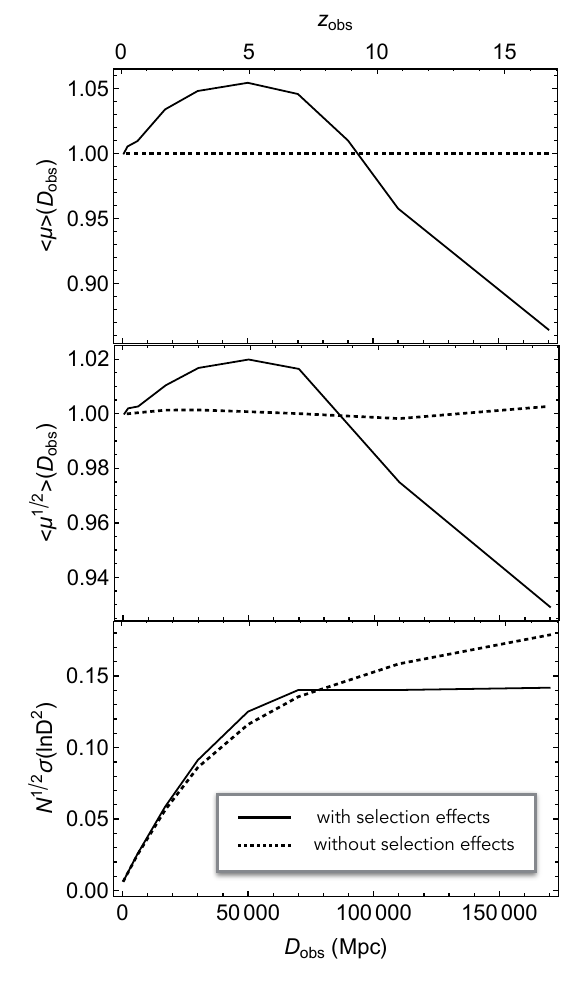}
\end{center}
\caption{\label{FinalPlotD}
Bias and uncertainty of the distance estimator $\hat{D}$ as a function of the observed luminosity distance (here $z_{\text{obs}}$ is derived from $D(z_{\text{obs}})=D_{\text{obs}}$ assuming our fiducial cosmology). We show both the bias on distance and on distance square, $\langle \mu^{1/2}\rangle$ and $\langle \mu\rangle$, respectively. 
We explicitly show the effect of including selection effects in the analysis.
}
\end{figure}

This last remark may be better understood by looking at Fig.\,\ref{FinalPlotD}, which shows the mean and variance of the distance estimator as a function of observed distance. The average magnification increases from 1 as we increase $D_{\text{obs}}$, it reaches a peak and then goes down to values $< 1$.
This can be explained by looking at the MBHB redshift distribution in Fig.~\ref{pop} (see also Fig.~3 in \citet{Klein:2015hvg}): being the distribution of sources peaked around redshift $8-10$, if a source is observed from high distance it is much more probable that this is actually a source at distance $D(z)<D_{\text{obs}}$, hence a de-magnified one.
 %it has been magnified  that  $D_{\text{obs}}=D(z)/\sqrt{\mu}$ 
On the other hand, a source at $z\lesssim 8-10$ has more chances to be a magnified one, as there are many sources at higher redshift with $D(z)>D_{\text{obs}}$.\\

\begin{figure}
\begin{center}
\includegraphics[width=0.39\textwidth]{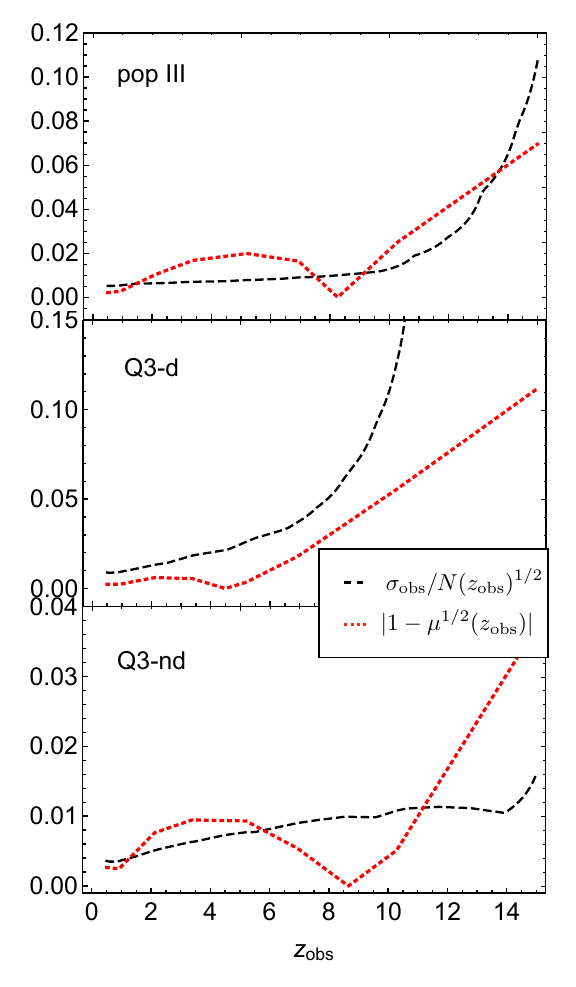}
\end{center}
\caption{\label{BiasDobs}
Comparison between bias and (relative) uncertainty $\sigma_{\text{obs}}=\sigma(\ln D)(z_{\text{obs}})$ on the distance estimator, as function of the observed redshift, for the three astrophysical models under study. The uncertainty at a given distance scales with the square root of the number density of detectable sources from that distance. We assumed a mission lifetime of 4 years.
}
\end{figure}

\section{Cosmological and astrophysical implications}

%\giulia{I would try to make this paragraph a bit less pessimistic. If I read now it seems that we are saying look what we do is useless. It is probably enough to emphasize the fact that there is an effect at high redshift, give astro implication and then say cosmology however is not affected}
As mentioned above, LISA MBHB standard sirens can be used to probe the expansion of the universe up to $z\sim 8$, above which EM counterparts are no longer expected to be observed \citep{Tamanini:2016uin}.
Since we found that selection effects mainly affect the MBHB distance measurements at $z\gtrsim 10$, we do not expect them to bias LISA standard siren analyses.
In fact, by using a similar procedure to the one outlined in \citet{Speri:2020hwc}, we checked that $\Lambda$CDM cosmological parameters will not be biased for any realistic number of MBHB standard sirens that LISA will observe.
Furthermore, by using eq.~\eqref{eq:error_fit} for LISA MBHB lensing uncertainty measurements, we find that forecasts on $\Lambda$CDM parameters are consistent with previous results.
Constraints on $\Lambda$CDM are only marginally better
%($\sim 6-7\%$ on average \nt{runs ongoing to update these estimates.. still running, should be over by Monday})
than the one reported in the literature \citep{Tamanini:2016uin,Belgacem:2019pkk,Speri:2020hwc} (differences are due to the slightly better fit provided by eq.~\eqref{eq:error_fit} with respect to the one of \citet{Hirata_2010}).
Selection effects could however affect other cosmological measurements based on weak lensing measurements with standard sirens, such as for example cross-correlations with EM matter surveys \citep{Congedo:2018wfn}.
The implications on such analyses are however more complicated to assess and their investigation will be left for future studies.

Important implications due to lensing selection effects will instead arise in MBHB population studies.
The redshift dependence of the MBHB merger rate yields information on the processes shaping the evolution and growth of the MBHB population, their environment and their relation with galaxy assembly (see e.g.~\citet{Haehnelt:1994wt,Kauffmann:1999ce,2003ApJ...582..559V,Enoki:2004ew,Sesana:2004gf,Micic:2007vd,Barausse:2012fy,Colpi:2014poa,Umeda:2016smj,2016MNRAS.457.3356V}).
Reconstructing the redshift and mass distribution of MBHB mergers constitutes thus an important scientific objective for the LISA mission \citep{Sesana:2007sh,Sesana:2010wy,Colpi:2019yzd}.
Such reconstructions will be performed by fitting merger rate distribution models to the MBHBs detected by LISA, in analogy to analyses for stellar-mass black hole binaries currently implemented with LIGO-Virgo observations \citep{Fishbach:2018edt}, which have already been shown to be affected by lensing selection effects; see e.g. \citet{Dai:2016igl,Oguri:2018muv}.
Since these analyses will require a sampling in redshift (and mass) of the LISA MBHB merger distribution, the effect of the bias on the distance estimator becomes in this case important. % when we consider the total population of sources (with or without counterpart).
The reason is that, although we have a high statistics of events and the bias on the distance estimator is independent on the sample size, the variance of the distance measurement associated to each bin decreases with $\sqrt{N}$.
In figure~\ref{BiasDobs} we compare bias and uncertainty for the sample of sources with SNR$>$8 (computed here neglecting lensing), as a function of the observed redshift (derived from the $D_{\rm obs}$ assuming our fiducial cosmology).
For this figure we present results for all three models considered in \citet{Klein:2015hvg}: popIII and the other two heavy-seed Q3 MBH population models (with and without delays between galaxy merger and MBHB merger, respectively ``Q3-d'' and ``Q3-nd''). 
We see that for popIII and Q3-nd, the bias is always bigger than the uncertainty at almost all observed redshift of interest. % for $z<8$.  
This implies that all analyses based on a sampling in $z_{\rm obs}$ or $D_{\rm obs}$ are expected to be strongly biased by lensing selection effects.
Further work is thus required to understand how to deal with this bias and be able to extract relevant astrophysical information from the LISA redshift distribution of MBHB mergers.
In the Q3-d scenario instead, lensing selection effects will have negligible implications on any population study as the bias will always be lower than the uncertainty of each redshift/distance bin.
This is due to the lower number of sources detected by LISA in the Q3-d model (cf.~\citet{Klein:2015hvg}) which implies that $\sigma_{\rm obs}/\sqrt{N(z_{\rm obs})}$ remains sufficiently big.
Note finally that these results are based on an observational time of 4 years, while a possible extension up to 10 years is envisaged for the LISA mission \citep{Audley:2017drz}.
For longer observational periods we expect the results above to show an even more marked bias due to lensing selection effects, as $N(z_{\rm obs})$ will on average grow linearly with the observational time.\\
%\nt{I added the last two sentences}\giulia{Good}\\

%\giulia{Added square root}\\

%\giulia{We need to better understand how to use this and which consequence to derive. I would also metion en passant that bias is significant when using BBH for cosmology with statistical redshift determination}
%\nt{I added some general comments on the astro implications. I don't think this will affect dark siren measurements, as no sampling in redshift is done with those analyses}
 
\section{Discussion and conclusion}

%\giulia{Try to say in a few words what selection effects are} 
We have presented a method to assess the effect of weak and strong lensing on the estimation of the luminosity distance for a population of astrophysical GW sources, taking into account selection effects due to the finite sensitivity of a GW detector.  
%With selection effects here we denote the following phenomena: given a sample of observed astrophysical sources, it is more probable that the magnified sources are those just below threshold (in the absence of magnification). Or, alternatively: 
Since a realistic GW detector has a finite horizon, the probability that a source detected at redshift $z$ is magnified is higher than the probability that it is de-magnified. As a consequence,  
the mean of the distribution of magnification for sources at sufficiently high redshift, is shifted from $1$ to higher values. These effects, which we dubbed \emph{lensing selection effects}, disappear in the limit of a perfect instrument (or analogously, in the limit of sources at redshift much lower than the instrument horizon).\footnote{We stress that our derivation of lensing selection effects relies on the geometric optics approximation. We expect wave effects to become non-negligible in the LISA waveband only when dealing with diffusion off sub-galactic structures, see e.g.\,\citet{Takahashi:2003ix,PhysRevLett.80.1138,Takahashi:2016jom, Dolan:2007ut, Cusin:2018avf,  Cusin:2019rmt, Dalang:2021qhu}. Diffraction on sub-galactic scales makes lenses on those scales effectively transparent to GW in the LISA band \citep{Takahashi:2003ix}, in contrast with what happens for lensing of EM sources (see e.g.\,\citet{Fleury_2015}) as the EM spectrum is at much
%higher frequency than any observable GW.}
lower wavelengths than any relevant astrophysical structure at cosmological scales. 
%nt{just rephrased a bit here to be clearer}
}

%A detailed inclusion of wave effects in our method will be presented in a future work. }
%\nt{good, but I wouldn't promise anything}\giulia{alright, no fake promise}

Although this fact was already pointed out in the literature, see e.g.~the discussion section of \citet{Hirata_2010}, this is the first place where a quantitative estimate of the selection effects on the determination of the luminosity distance of MBHB has been presented. In particular we provide an unbiased estimator for the luminosity distance as a function of redshift, which includes the effect of lensing magnification. We explicitly studied the case of a population of MBHB mergers visible by LISA.
Fixing the reference astrophysical model to the popIII source distribution of \citet{Barausse:2012fy,Klein:2015hvg}, we computed the bias and the variance of the distance for the entire population (with or without EM counterpart). 

%We stress that these results depend on the astrophysical model for source distribution.
%\giulia{I would remove this paragrph. See what i added below. When studying lensing of a given population of sources, one should apply the method presented here and compute the variance and mean of the distance estimator using the distribution of sources under study. However, the general picture we find is universal: the mean magnification grows from 1 to higher values as we move up in redshift and the uncertainty induced by lensing on measurement of distance increases faster with redshift as we exit the instrument horizon.  We stress that while this uncertainty as function of redshift decreases with the number density of detectable sources at that redshift, the bias is \emph{irreducible} being independent of the sample size.}
%\nt{fine by me}

While the effect of the bias on the distance estimator for sources with EM counterpart is typically below the variance threshold, it may become relevant for high redshift sources when the statistics of detectable events is large. If no EM counterpart is present, a more useful quantity to look at is the distribution of magnification as function of observed distance.  We compare the mean magnification to the variance as function of observed distance in Fig.\,\ref{BiasDobs}. This comparison shows that sources observed from high distances, specifically beyond the peak of the merger distribution, are on average de-magnified and that the average magnification is typically larger than the associated variance, hence it should not be neglected when working out distance estimates  with a statistical approach, such as for example in source population studies and in particular to determine the merger rate as a function of redshift. 

Notice however that our estimate of the variance of luminosity distance used the Cramer-Rao bound, as explained in Sec.~\ref{estimator}. A more realistic analysis would estimate the variance of luminosity distance as the second moment of eq.~(\ref{defD}), re-running a ray-tracing simulation taking into account wave effects in the PDF of lensing magnification. For this reason the value of redshift at which bias and variance are comparable in value may be understimated in our analysis. Notice however that the scope of this paper is to point out the methodology to follow in future realistic lensing studies and not to provide precise and conclusive estimates. Moreover one of the main results of our preliminary analysis is that for standard sirens the lensing bias is always within the variance level. This is a conservative result as we use the Cramer-Rao bound to compute the variance of luminosity distance and a more realistic analysis would typically give a an estimated value of the variance above the Cramer-Rao bound. In other words, a more realistic study would find that for standard sirens lensing bias is still be within the estimated variance level.

As mentioned above, we focused our investigations on the popIII MBHB population model considered in \citet{Barausse:2012fy,Klein:2015hvg}.
We made this choice because the popIII model is the one for which the implications of the selection effects are higher.
For the other heavy-seed MBHB population models of \citet{Barausse:2012fy,Klein:2015hvg}, lensing selection effects  are less relevant since their redshift distributions are shifted towards lower redshift values and their mass distribution towards higher masses.
Because of this almost all heavy-seed MBHB merger events are detectable by LISA \citep{Klein:2015hvg}, and a few sub-threshold events have the probability of being magnified enough to result detectable. 
We also showed that only for the heavy seed Q3-nd model of \citet{Barausse:2012fy,Klein:2015hvg} population studies will be biased by selection effects, while for the Q3-d model no such bias should arise (cf.~Fig.~\ref{BiasDobs}).
As explained above, this is due to the higher number of LISA detection in Q3-nd with respect to Q3-d which drives the distance uncertainty on the luminosity distance estimator given by eq.~\eqref{sigma} to lower values.
Additionally we remark that our result for the distribution of magnification are taken from the ray-tracing simulation of \citet{Takahashi:2011qd}, which does not include the effect of baryons. The inclusion of baryons typically enhances the high-magnification tail of the distribution.
Our results of lensing selection effects should therefore be taken as conservative estimates.

For the benefit of future investigations, we also supplied here simple analytical expressions for the lensing distance error containing for the first time the contribution due to selection effects.
These ready-to-use expressions can be employed in LISA MBHB analyses requiring realistic distance determination of each source. 
%Although as we showed, lensing selection effects appear to be irrelevant at the redshift range over which EM counterpart to MBHB mergers are expected to be observed, the clearest example of such analyses is provided by cosmological studies with standard sirens, which will need to take into account the effect of lensing on the luminosity distance errors as this is on average always dominant with respect to LISA instrument uncertainties at high redshift.
The clearest example of such analyses is provided by cosmological studies with standard sirens, which will need to take into account the effect of lensing on the luminosity distance errors as this is on average always dominant with respect to LISA instrument uncertainties at high redshift, even if, as we showed, lensing selection effects appear to be irrelevant at the redshift range over which EM counterpart to MBHB mergers are expected to be observed.

%\giulia{I added here this paragraph}
In this work, we assumed to know the underlying distribution of sources, and we estimated the bias in the reconstruction of distance distribution, induced by lensing selection effects. Of course, in realistic population studies, one rather wants to infer the (unknown) properties of a given source population, starting from an observed sample of sources. To be concrete, let us assume that one wants to extract the merger rate per unit chirp mass and cosmological redshift, from the observed rate. Then one should use that the two distributions are related by (fixing a reference observation time)
\begin{align}\label{popstudy}
&\frac{d^2N(\mathcal{M}_{\text{obs}}, z_{\text{obs}})}{d\mathcal{M}_{\text{obs}}dz_{\text{obs}}}=\nn\\
&=\int d\ln\mu\frac{d\mathcal{P}_z(\mu, z)}{d\ln \mu}\frac{d^2N(\mathcal{M}, z)}{d\mathcal{M}dz}\vline\frac{\partial(\mathcal{M}, z)}{\partial(\mathcal{M}_{\text{obs}}, z_{\text{obs}})}\vline\,,
\end{align}
where $\mathcal{M}_{\text{obs}}(1+z_{\text{obs}})=\mathcal{M}(1+z)$ and the Jacobian (at fixed $\mu$) is given by
\be
\vline\frac{\partial(\mathcal{M}, z)}{\partial(\mathcal{M}_{\text{obs}}, z_{\text{obs}})}\vline=\frac{(1+z_{\text{obs}})}{(1+z)}\frac{D'(z_{\text{obs}})}{D'(z)}\frac{D(z)}{D(z_{\text{obs}})}\,.
\ee
We observe that in eq.\,(\ref{popstudy}),  $\mathcal{P}_z(\mu, z)$ is the observed distribution of magnification, with selection effects included (cf.~eq.\,(\ref{PDistr})), which in turn depends on mass and redshift distribution of sources. In analyses with real observational data, it is the right hand side of eq.~(\ref{popstudy}), considered as function of source-frame chirp mass and cosmological redshift, that has to be fitted to the observed rate to extract information on the intrinsic redshift and mass distribution of the population under study.
Studies along these lines for the case of ground-based detectors have been considered for example by \citet{Dai:2016igl, Oguri:2018muv}.
%\giulia{Sei d'accordo con quello che scrivo?}\nt{si, ho solo rimaneggiato leggermente. forse meglio citare Dai+ come esempio qui?}\giulia{We can. But notice that the analog expression in Dai+ uses a probability distribution without selection effects (not sure I understand why)...We could say something generic as. A study along these lines for the case of ground-based detector has been recently presented in \citet{Dai:2016igl, Oguri:2018muv}}. \nt{si cosi \'ebbuono}

To conclude, lensing of high-redshift GW sources biases the observed GW signal, hence contaminating the reconstruction of the astrophysical properties of the population of emitting sources (above of all distances and source-frame chirp masses).
For high-redshift GW sources, the characterisation of all the implications due to lensing, including selection effects due to the specifics of a given instrument, is thus of paramount importance, not only to infer in an accurate way their astrophysical properties across cosmic time, but also to be able to use high-redshift GW sources as a new compelling cosmological probe.\\

%\nt{did we add the citation requests that we received in the end?}\giulia{I have cited \citet{2019MNRAS.486.2336D} in the list of better catalogues.  Suvdip's requests i don't know. Maybe we can cite the first of his list.} \nt{your call. just let me know so that I can reply to the emails}\\

%\giulia{slightly shorten the last sentence. Old version commented}\\
%The characterisation of all the implications due to lensing, including selection effects due to the specifics of a given instrument, is thus of paramount importance to accurately determine distances from high-redshift GW sources ; this is relevant to infer in an accurate way their astrophysical properties across cosmic time but also to be able to use them as new compelling cosmological probes.\\

\noindent 
\textit{Acknowledgements.}
We are very grateful to Ruth Durrer and Irina Dvorkin for valuable discussions during different stages of this project. We also thank Enrico Barausse, Enis Belgacem, Giuseppe Congedo, Daniel Holz, Francesca Lepori and Jean-Philippe Uzan for discussions and comments on the manuscript.  
GC thanks Alessandra Buonanno and the entire Astrophysical and Cosmological Relativity division at the Max Planck Institute for Gravitational Physics (Albert Einstein Institute) in Potsdam for their warm hospitality during the first stage of this work. The work of GC is supported by the Swiss National Science Foundation.\\

\noindent
\textit{Data Availability Statement.}
The data underlying this article will be shared on reasonable request to the corresponding author.

\appendix

\section{Bounding the variance with the Fisher information}

Let us assume to have an unknown deterministic parameter $x$ to be estimated from $N$ independent observations of $X$, each distributed following a probability density function $f(x, X)$. The variance of any unbiased estimator $\hat{x}$ of $x$ is bounded by the reciprocal of the Fisher information 
\be\text{var}(\hat{x})\geq \frac{1}{I(x)}\,,
\ee
where
\be
I(x)=N\int dx f(x, X)\left(\frac{d}{dx}f(x, X)\right)^2\,.
\ee
If we have a biased estimator $\hat{T}$ whose expectation value is not $x$ but a finite function $\psi(x)$ of it, the bound is
\be\label{Cramer}
\text{var}(\hat{T})\geq \frac{|\psi'(x)|^2}{I(x)}\,,
\ee
where $\psi'(x)=\partial \psi/\partial x$. Let us use this result in our context. Let us fix a given redshift $z$ and  assume that we have $N(z)$ measurements of luminosity distance from a given bin around $z$.  We pose $x=\ln \mu$ and $\hat{T}=\hat{D}^2$ with $\hat{D}^2\equiv \mu\bar{D}^2_{\text{obs}}=e^x \bar{D}^2_{\text{obs}}$. Then using (\ref{Cramer}), we get the following upper bound on the variance of the distance estimator 
\be
\text{var}(\hat{D}^2)\geq \frac{\langle \hat{D}^2\rangle^2}{I(x)}\,,
\ee
where
\be
I(x)=N(z)\int dx \mathcal{P}_z(x, z)\left(\frac{d}{dx}\mathcal{P}_z(x, z)\right)^2\,,
\ee
and the dependence on $z$ on the left hand side is understood. 
Following the same reasoning, but considering $\hat{T}=\ln\hat{D}^2\equiv \ln\bar{D}^2_{\text{obs}}+x$, we have 
\be
\text{var}(\ln\hat{D}^2)\geq \frac{1}{I(x)}\,. 
\ee

\bibliographystyle{mn2e}
\bibliography{refs}

\end{document}